%% file: main.tex
\documentclass[journal,twoside,web]{ieeecolor}
\usepackage{generic}
\usepackage{cite}
\usepackage{amsmath,amssymb,amsfonts}
\usepackage{algorithmic}
\usepackage{graphicx}
\usepackage{algorithm,algorithmic}
\usepackage{hyperref}
\hypersetup{hidelinks}
\usepackage{booktabs}
\usepackage{textcomp}
\usepackage{comment}

\usepackage{siunitx} 
\usepackage[acronym,nopostdot,nohypertypes=acronym]{glossaries}
\input{acronyms}

\newcommand{\glsexplain}[1]{\acrshort{#1}, \acrlong{#1}}

\def\BibTeX{{\rm B\kern-.05em{\sc i\kern-.025em b}\kern-.08em
    T\kern-.1667em\lower.7ex\hbox{E}\kern-.125emX}}
\begin{document}
\title{Building Privacy-and-Security-Focused Federated Learning Infrastructure for Global Multi-Centre Healthcare Research}
\author{Fan~Zhang,
        Daniel~Kreuter,
        Javier~Fernandez-Marques,
        BloodCounts!~Consortium,
        Gregory Verghese,
        Bernard Butler,
        Nicholas~Lane,
        Suthesh~Sivapalaratnam,
        Joseph~Taylor,
        Norbert~C.J.~de~Wit,
        Nicholas~S~Gleadall,
        Carola-Bibiane~Schönlieb,
        and~Michael~Roberts
\thanks{F. Zhang, D. Kreuter, C.-B. Schönlieb, and M. Roberts are with the Department of Applied Mathematics and Theoretical Physics, University of Cambridge, Cambridge, UK. D. Kreuter is also with the Precision Health University Research Institute, Queen Mary University of London, London, UK. M. Roberts is also with the Department of Medicine, University of Cambridge, Cambridge, UK. Corresponding author: Michael Roberts (e-mail: mr808@cam.ac.uk)}
\thanks{J. Fernandez-Marques and N. Lane are with Flower Labs, Cambridge, UK. N. Lane is also with the Department of Computer Science and Technology, University of Cambridge, Cambridge, UK.}
\thanks{G. Verghese is with PharosAI, London, UK, and Cancer Bioinformatics, School of Cancer \& Pharmaceutical Sciences, Faculty of Life Sciences and Medicine, King’s College London, UK.}
\thanks{B. Butler is with the Department of Computing and Mathematics, South East Technological University, Waterford, Ireland.}
\thanks{J. Taylor and S. Sivapalaratnam are with the Precision Health University Research Institute, Queen Mary University of London, London, UK, and with the Department of Clinical Haematology, Royal London Hospital, Barts Health NHS Trust, London, UK. S. Sivapalaratnam is also with the Blizard Institute, Queen Mary University of London, London, UK.}
\thanks{N. C. J. de Wit is with the Maastricht University Medical Centre, Maastricht, Netherlands.}
\thanks{N. S. Gleadall is with the Victor Phillip Dahdaleh Heart and Lung Research Institute; the Department of Haematology, University of Cambridge; and NHS Blood and Transplant, Cambridge Biomedical Campus, Cambridge, UK.}
\thanks{The BloodCounts! Consortium: A list of authors and their affiliations appears at the end of the paper.}}

\maketitle

\begin{abstract}
Collaborative healthcare research across multiple institutions increasingly requires diverse clinical datasets, but cross-border data sharing is strictly constrained by privacy regulations. \Gls{fl} enables model training while keeping data local; however, existing frameworks largely remain proof-of-concept, inadequately addressing privacy risks arising from unauthorised participation, misuse, and unaccountable computation, with insufficient enforceable governance mechanisms (\gls{aaa}) required for real-world clinical deployment. This paper presents FLA\textsuperscript{3} (\gls{fl} with \gls{aaa}), a governance-aware \gls{fl} platform 
that operationalises regulatory obligations through runtime policy enforcement. FLA\textsuperscript{3} integrates \gls{xacml}-compliant attribute-based 
access control, cryptographic accounting, and study-scoped federation directly 
into the \gls{fl} orchestration layer to enforce institutional sovereignty and protocol adherence. We evaluate FLA\textsuperscript{3} through two complementary studies. First, we demonstrate operational feasibility by deploying the platform infrastructure across five \textit{BloodCounts!} Consortium institutions in four countries (United Kingdom, Netherlands, India, and The Gambia), confirming that governance enforcement operates correctly under realistic network and regulatory constraints. Second, we assess clinical utility through simulated federation of full blood count data from 54,446 samples from 35,315 subjects across 25 centres in the INTERVAL study, demonstrating that FLA\textsuperscript{3} achieves predictive performance comparable to centralised training while strictly enforcing governance constraints. These results confirm that enforceable governance, when treated as a first-class privacy-preserving control, is feasible and enhances trustworthiness for scalable \gls{ai} in healthcare deployments requiring cross-jurisdictional compliance.
\end{abstract}

\begin{IEEEkeywords}
Federated Learning, Healthcare \gls{ai}, Privacy-Preserving Systems, Data Governance, Access Control, \gls{xacml}, Auditability.
\end{IEEEkeywords}

\glsresetall

\section{Introduction}
\label{sec:introduction}

\IEEEPARstart{M}{ulti-centre} healthcare research increasingly depends on data-intensive \gls{ai} methods that require large, diverse, and representative clinical datasets. However, the secondary use of patient-level healthcare data across institutional and national boundaries remains tightly constrained by privacy regulations such as the \gls{hipaa} in the United States~\cite{HIPAA1996}, the \gls{gdpr} in the European Union~\cite{EU2016_679}, and the UK GDPR in the United Kingdom~\cite{UKGDPR2016}. These statutory frameworks, together with institutional governance policies, restrict or prohibit centralised aggregation of health data, particularly in cross-border collaborations. Consequently, many clinically relevant \gls{ai} models are never trained on sufficiently large datasets to realise their full capacity or remain infeasible despite the existence of suitable data distributed across multiple institutions.

\Gls{fl} has become a critical privacy-preserving paradigm for healthcare \gls{ai}, enabling collaborative model training without the exchange of sensitive patient data~\cite{mcmahanCommunicationEfficientLearningDeep2023}. However, privacy risks in regulated healthcare settings extend beyond information leakage through data transfer or model updates. Under clinical research governance and data protection regulations such as \gls{gdpr} and \gls{hipaa}, computation performed outside the approved institution, after approval expiry, or for unapproved purposes constitutes unauthorised processing of personal data, even when raw data remain local. 

Prior studies have demonstrated the technical feasibility of \gls{fl} across medical imaging~\cite{shellerFederatedLearningMedicine2020}, computational pathology~\cite{ogierduterrailFederatedLearningPredicting2023,luFederatedLearningComputational2022}, electronic health records~\cite{choudhuryDifferentialPrivacyenabledFederated2020}, and clinical prediction tasks~\cite{adnanFederatedLearningDifferential2022}. However, systematic reviews report that only approximately 5\% of \gls{fl} studies in healthcare involve real-world deployment, with the majority remaining proof-of-concept evaluations that assume trusted participants and idealised operating conditions~\cite{teoFederatedMachineLearning2024}. Our recent systematic review of 89 \gls{fl} healthcare methodologies revealed pervasive governance deficiencies: 87/89 (98\%) lacked node authentication and critically, none provided a peer-reviewed, openly specified governance implementation compliant with institutional governance requirements~\cite{zhangRecentMethodologicalAdvances2024}. Subsequent analysis identified that governance research itself is equally underdeveloped, with 
only 7 of 39 reviewed papers examining \gls{fl} governance mechanisms and none providing operational implementation guidance~\cite{edenScopingReviewGovernance2025}. In regulated healthcare environments, however, trusted participants and secure operational contexts cannot be assumed. 
Therefore, compliance, governance, and operational accountability are primary system requirements rather than secondary considerations, a position reinforced by the widespread adoption of Trusted Research 
Environments (TREs)~\cite{SATRE_specification_2025,HDRUK_TREs_2026} across 
the United Kingdom and the European Union, whose architectural constraints on data egress, software installation, and network access impose additional 
governance obligations that existing \gls{fl} frameworks do not address.

The \gls{fl} workflow itself constitutes a shared operational system requiring stringent \gls{aaa} controls. For example, if an ethics approval expires mid-study: without runtime enforcement, the associated node continues contributing to model training undetected, rendering the entire study non-compliant and potentially invalidating all results. 

Existing open-source \gls{fl} frameworks provide foundational security mechanisms
but lack integrated enforcement of healthcare-specific governance requirements.
Flower~\cite{beutelFlowerFriendlyFederated2022} implements TLS-based communications, node
authentication and \gls{cli} user authentication and authorisation, but does not enforce
study-scoped authorisation or temporal validity constraints at the orchestration
layer. PySyft~\cite{zillerPySyftLibraryEasy2021} focuses on cryptographic privacy
mechanisms (secure aggregation, differential privacy) for data protection rather
than \gls{fl} workflow governance. NVIDIA FLARE~\cite{rothNVIDIAFLAREFederated2022}
provides role-based authorisation policies and site-level access control, but
authorisation is configured at provisioning time and does not support runtime
enforcement of study-specific participation constraints, automatic authorisation
expiry, or temporal validity predicates. Consequently, these platforms lack
native mechanisms for the continuous, policy-driven enforcement of study-scoped
authorisation, time-bounded participation, and fine-grained access control
required for healthcare deployment. This gap introduces regulatory compliance risks: without runtime 
enforcement, institutions cannot ensure that computation occurs only within 
approved study boundaries, temporal validity windows, or authorised participant 
sets, potentially rendering entire studies non-compliant with ethics approvals 
and data protection regulations. To address these governance gaps, we systematically analyse healthcare data protection regulations including 
multiple legal frameworks such as \gls{gdpr} (European Union and United Kingdom), 
\gls{hipaa} (United States), \gls{dpdpa} (India), and \gls{ecowas} (West 
Africa) policy, to identify common governance obligations, which we formalise as five 
enforceable requirements (R1--R5) for healthcare \gls{fl} systems. This cross-jurisdictional 
analysis identifies common governance obligations that must be satisfied for 
compliant healthcare \gls{fl} deployment. Privacy-enhancing techniques such as differential 
privacy~\cite{dworkAlgorithmicFoundationsDifferential2014} for gradient 
perturbation and secure multi-party 
computation~\cite{Dong2021MPCHealthcare} for encrypted aggregation mitigate 
information leakage during model training. However, they do not inherently enforce institutional accountability, time bounded authorisation, or study-specific compliance. These requirements must be implemented through complementary governance mechanisms operating at the system orchestration layer.

These challenges are further amplified in global, multi-centre collaborations. Participating institutions may exhibit substantial heterogeneity in governance requirements, security postures, network configurations, and patient population characteristics that drive statistical heterogeneity in clinical data~\cite{edenScopingReviewGovernance2025}. Many healthcare organisations operate within restrictive network environments that prohibit inbound connectivity, complicating server-initiated communication. In addition, clinical data are implicitly \gls{noniid} due to demographic, 
geographic, and socioeconomic variation across patient populations, as well as site-specific laboratory practices~\cite{chenEffectiveNonIIDDegree2025}, requiring aggregation strategies that can accommodate heterogeneous data distributions while maintaining model utility. Deployable \gls{fl} systems must therefore address governance enforcement and statistical heterogeneity under operational constraints imposed by network policies, security requirements, and \gls{noniid} data.

We address these gaps through FLA\textsuperscript{3} (\gls{fl} 
with \gls{aaa}), a governance-aware \gls{fl} platform 
that integrates \gls{xacml}~\cite{OASIS_XACMLv3_2013} compliant policy enforcement, cryptographic audit mechanisms, 
and study scoped federation directly into \gls{fl} orchestration. We validate 
FLA\textsuperscript{3} through operational deployment with the 
BloodCounts! Consortium, an international haematology research collaboration 
including five healthcare institutions across four countries (United Kingdom, Netherlands, India, and The Gambia). This deployment context exemplifies the governance challenges that motivated our work. Specifically: (a) institutions operate under different regulatory frameworks (\gls{gdpr}, \gls{dpdpa}, \gls{ecowas}), (b) sites exhibit heterogeneous network security configurations (including egress-only connectivity, restrictions on software installation, and data transfer controls) and clinical data distributions, and (c) all require study-specific approvals with temporal validity constraints. BloodCounts! focuses on the analysis of \gls{fbc} data, where \gls{fbc} measurements exhibit substantial cross-population variability driven by genetic, environmental, and socioeconomic factors~\cite{ambayyaHaematologicalReferenceIntervals2014}, limiting the generalisability of models trained on single-site data. The haematology use case demonstrates the FLA\textsuperscript{3}'s clinical utility, and the architectural decisions address governance requirements common across healthcare \gls{fl} deployments. The applicability of FLA\textsuperscript{3} to healthcare data is demonstrated using simulated federation of data from the INTERVAL study to enable controlled multi-centre validation.

In this work, we present an integration of \gls{xacml} compliant attribute-based access control~\cite{OASIS_XACMLv3_2013} with \gls{fl} orchestration for healthcare \gls{ai}. We build upon Flower~\cite{beutelFlowerFriendlyFederated2022}, an open-source \gls{fl} framework, extending it with policy-driven governance controls. We address the gap between \gls{fl} research prototypes and deployable healthcare systems by designing and implementing policy-driven governance, enforceable access control, auditable accountability, and federated personalisation within a single production-ready platform. Our contributions are as follows:

\begin{enumerate}
    \item \textbf{Regulatory-Derived Governance Requirements:} We systematically analyse cross-jurisdictional healthcare data protection regulations (\gls{hipaa}, \gls{gdpr}, \gls{dpdpa}, \gls{ecowas}) and clinical research governance frameworks, identifying common obligations and formalising them as enforceable system properties (R1--R5 in Section~\ref{sec:governance-threat-model}).

    \item \textbf{Policy-Driven \gls{aaa} Framework:} We design \gls{xacml}-compliant authorisation and cryptographic accounting mechanisms integrated with Flower's authentication. The framework enforces fail-closed evaluation (denying access when policy evaluation fails or attributes are missing) at all \gls{fl} lifecycle points and generates cryptographically signed audit records to support regulatory audit and institutional accountability.

    \item \textbf{Multi-Study Federation:} We implement study-scoped federation in which each research study operates as an independent collaborative unit with its own participant set, authorisation policies, and temporal validity constraints, enabling a single platform to support multiple concurrent clinical studies.
    
    \item \textbf{Governance-Preserving Personalisation:} We demonstrate that policy-driven governance enforcement does not degrade personalised FL performance when integrated with FedMAP~\cite{zhangFedMAPPersonalisedFederated2025}. Evaluation using data from 25 centres participating in the INTERVAL study~\cite{diangelantonioEfficiencySafetyVarying2017} shows federated training significantly improves performance compared with individual training while achieving predictive performance comparable to centralised training.

    \item \textbf{Open-Source Reference Implementation:} We release a complete, open-source implementation (\url{https://github.com/bloodcounts/FLAAA}), providing a reference architecture for governance-aware \gls{fl} deployment in regulated healthcare environments.
    
\end{enumerate}

\section{Related Work}
\label{sec:related}

This section reviews prior work considering healthcare, privacy and security mechanisms in \gls{fl}, governance and access control frameworks, and deployment-oriented \gls{fl} systems.

\subsection{FL in Healthcare and Privacy Preservation}
Prior \gls{fl} studies in healthcare have focused primarily on validating model performance under data locality constraints (where data remain at source institutions and are not transferred)~\cite{shellerFederatedLearningMedicine2020,choudhuryDifferentialPrivacyenabledFederated2020,adnanFederatedLearningDifferential2022}. However, maintaining data locality introduces a challenge: healthcare data exhibit substantial statistical heterogeneity due to demographic, geographic, and institutional variation across patient populations, as well as differences in clinical practices and measurement protocols. Personalised \gls{fl} approaches address this by combining global knowledge sharing with local adaptation~\cite{yePersonalizedFederatedLearning2025}. Examples of these include methods such as FedProx~\cite{li2020federatedoptimizationheterogeneousnetworks} which uses proximal regularisation (adding penalty terms that constrain local updates to remain close to the global model), and FedPer~\cite{arivazhagan2019federatedlearningpersonalizationlayers} which separates models into shared base layers and site-specific head layers. We previously developed FedMAP~\cite{zhangFedMAPPersonalisedFederated2025}, a personalised \gls{fl} method based on local \gls{map} estimation with a learned prior parameterised by an input-convex neural network (a neural network architecture that is convex with respect to its inputs), enabling sites with \gls{noniid} data to maintain local accuracy and leveraging shared knowledge from other institutions. 

However, as reviewed in Section~\ref{sec:introduction}, systematic analyses reveal that the majority of healthcare \gls{fl} studies remain proof-of-concept deployments that do not address governance requirements~\cite{teoFederatedMachineLearning2024,zhangRecentMethodologicalAdvances2024}. This work demonstrates that enforceable \gls{aaa} controls are compatible with \gls{fl} aggregation strategies by integrating them with FedMAP, a personalised \gls{fl} method, though the governance framework applies to any \gls{fl} approach requiring institutional accountability and protocol compliance in deployed settings.

\subsection{Governance, Access Control, and Deployment in FL}

Governance research in \gls{fl} has identified organisational challenges 
in multi-institutional collaborations~\cite{edenScopingReviewGovernance2025, 
barbereauGovernanceFederatedLearning2025}, typically proposing conceptual 
frameworks based on data sharing agreements, ethics approvals, consortium 
structures, or commercial contractual agreements. The last of these is 
increasingly relevant as industry partners seek access to multi-centre 
platforms: enforcing strict study-level segregation is a prerequisite 
for protecting the intellectual property of each participating organisation 
and ensuring that proprietary model outputs or data-derived insights from 
one project cannot be accessed by participants of another. Access control models such as \gls{rbac} and \gls{abac} have been extensively studied in distributed systems and healthcare IT~\cite{centonzeRoleAttributeBasedEncryptionRABE2016, mukherjeeAttributeBasedAccess2017}, with \gls{xacml} providing standardised policy evaluation~\cite{OASIS_XACMLv3_2013}. However, these approaches operate primarily at the organisational level and lack runtime enforcement mechanisms within \gls{fl} systems, leaving privacy risks from unauthorised processing and accountability gaps unmitigated even when raw data remain local.

\gls{fl} frameworks such as Flower~\cite{beutelFlowerFriendlyFederated2022}, PySyft~\cite{zillerPySyftLibraryEasy2021}, NVIDIA FLARE~\cite{rothNVIDIAFLAREFederated2022}, and TensorFlow Federated~\cite{tensorflow_federated} provide basic authentication and limited authorisation mechanisms; however, these controls are typically coarse-grained, statically configured, and lack integrated, runtime-enforceable support for study-scoped, time-bounded governance and comprehensive auditing. Deployed healthcare systems including FeTS~\cite{shellerMultiinstitutionalDeepLearning2019}, MELLODDY~\cite{heyndrickxMELLODDYCrosspharmaFederated2024}, and Substra~\cite{galtierSubstraFrameworkPrivacypreserving2019} demonstrate practical feasibility but enforce governance through organisational mechanisms rather than continuous runtime evaluation. This approach cannot prevent computation outside approved boundaries once credentials are provisioned, creating compliance risks when approvals expire or participants exceed their authorised roles. This 
limitation extends to commercially deployed platforms such as Rhino Federated 
Computing~\cite{rhinofcp_platform}, which provide real-world multi-institution 
deployments but whose governance architectures remain proprietary and have not 
been described in peer-reviewed publications, precluding independent verification 
of runtime enforcement mechanisms. Our work addresses this gap by integrating \gls{xacml}-based authorisation and cryptographic auditing into the \gls{fl} orchestration layer, forming a privacy-preserving control plane that enforces approved participation and accountable execution at runtime, complementing organisational governance. The governance architecture is framework-agnostic, applicable to any \gls{fl} system with defined orchestration points for policy enforcement. Our reference implementation extends Flower~\cite{beutelFlowerFriendlyFederated2022}, building upon its \gls{mtls} authentication and lifecycle management to demonstrate integration with an existing deployed \gls{fl} runtime.

\section{Privacy and Governance Requirements and Threat Model}
\label{sec:governance-threat-model}

This section derives the privacy and governance requirements that our \gls{fl} framework must satisfy by analysing healthcare-specific regulatory and clinical research governance across our deployment jurisdictions, and by characterising the threat model unique to multi-institutional healthcare collaborations.

\subsection{Healthcare Governance Requirements}

Multi-centre healthcare research is governed by clinical research oversight 
frameworks including institutional review boards, medical research ethics 
committees~\cite{gradyInstitutionalReviewBoards2015}, and Data Access 
Committees (DACs)~\cite{cheahDataAccessCommittees2020}, which provide independent oversight of 
data release decisions and impose binding conditions on approved use. These regimes impose binding controls over who may participate, for what approved purpose, under which role and time constraints, and with what level of accountability. Such controls are formalised through ethics approvals, study protocols, and \glspl{dsa}~\cite{scheibnerDataProtectionEthics2020}. Non-compliance can invalidate studies, trigger regulatory action, and undermine patient trust, making these constraints enforceable requirements rather than static preconditions throughout the operational lifecycle of an \gls{fl} study.

Across healthcare collaborations, recurring governance expectations include institutional identity verification, study-specific authorisation, role based access control, temporal validity, and comprehensive auditability. These are formalised as requirements R1--R5 below, corresponding to \gls{aaa}.

Our current deployments span four settings: the United Kingdom and the Netherlands (\gls{gdpr}-based regimes complemented by strong clinical research governance), India (a consent-centric framework under the Digital Personal Data Protection Act (DPDPA)~\cite{DPDPAct2023}), and The Gambia (\gls{ecowas} aligned regulation with data protection policy 
framework~\cite{Gambia_DPP_2019}). Despite differing legal instruments, all require verified institutional participation, study-specific approval, and demonstrable accountability for computation involving patient data.

Clinical research in these jurisdictions is additionally governed by healthcare specific regulations that extend beyond general data protection. The \gls{hra}~\cite{UKPolicyFrameworkHSCResearch} coordinates Research Ethics Committee (REC) approval under the UK Policy Framework for Health and Social Care Research~\cite{UKPolicyFrameworkHSCResearch}, the EU Clinical Trials Regulation (536/2014)~\cite{EUClinicalTrialsReg5362014} mandates sponsor accountability in EU Member States, ICH Good Clinical Practice (E6~R2)~\cite{ICH_E6R2_Addendum} defines international standards for investigator accountability and audit trails, and India’s ICMR National Ethical Guidelines~\cite{ICMR_Guidelines_2017} establish requirements for institutional ethics oversight. These frameworks operationalise the governance expectations identified above (institutional identity verification, study-specific authorisation, and comprehensive audit trails), but existing \gls{fl} systems do not enforce them without explicit governance mechanisms, motivating our \gls{aaa} framework.

\subsection{Derived Governance Requirements}

From the cross-jurisdictional healthcare governance analysis above, we derive five governance requirements (R1--R5). These requirements capture the essential authentication, authorisation, and accountability constraints that appear consistently across all examined frameworks (GDPR, HIPAA, DPDPA, ECOWAS) and are enforceable through system-level controls. Each corresponds to a recurring obligation in clinical research governance and is expressed as an enforceable system property. The threat model formalised in Section~\ref{subsec:threat-model} identifies the specific attacks that these requirements mitigate.

R1: Authenticated Institutional Participation
All \gls{fl} participants must be verifiably identified at the institutional level and associated with recognised governance attributes (e.g., ethics approval identifiers, \gls{dsa} status). Clinical approvals and accountability in regulated healthcare research are issued to institutions rather than individuals, as established by \gls{hra} approval processes binding authorisation to named institutional sponsors~\cite{UKPolicyFrameworkHSCResearch} and ICH GCP requirements for investigator site qualification~\cite{ICH_E6R2_Addendum}. Data protection regulations reinforce this through \gls{gdpr} Art.~5(2) (controller accountability) and \gls{hipaa} \S164.312(a) and \S164.308(a)(4) (access control and traceability for systems processing protected health information).

R2: Study-Scoped Authorisation
Participation must be explicitly authorised for each study; approval for one protocol does not extend to others. Computation outside approved protocol scope constitutes a governance breach. Healthcare research frameworks mandate protocol-specific approval: \gls{hra}/REC~\cite{UKPolicyFrameworkHSCResearch}, EU Clinical Trials Regulation (EU CTR)~\cite{EU_CTR_536_2014}, and ICMR ethics committees~\cite{ICMR_Guidelines_2017} issue non-transferable, per-protocol authorisations. Data protection law provides the legal boundary for enforcement through \gls{gdpr} Art.~5(1)(b) (purpose limitation) and \gls{hipaa} provisions governing authorised research use or IRB waiver.

R3: Role-Based and Least-Privilege Access
Participating nodes must be constrained to protocol-approved roles (e.g., participant, observer) with no implicit privilege escalation at the federation level. Role-based access is specified by study protocols and governance agreements determining which institutions contribute model updates (participants) versus receive aggregated results only (observers), supported by ICH GCP delegation requirements~\cite{ICH_E6R2_Addendum} and Caldicott Principles 3--4~\cite{Caldicott_Principles_2021}. Within institutions, individual access to local \gls{fl} components and patient data is governed by \gls{hipaa} \S164.312(a) and \S164.308(a)(4).

R4: Temporal Validity
Authorisation must be time-bounded and automatically expire when ethics approvals lapse, protocols conclude, or \glspl{dsa} terminate. This operationalises \gls{gdpr} requirements for purpose limitation and storage limitation (Art.~5(1)(b), (e)) by enforcing time bounds aligned with healthcare research approvals, including \gls{hra}/REC validity periods~\cite{UKPolicyFrameworkHSCResearch}, EU clinical trials end dates~\cite{EU_CTR_536_2014}, and contractual \gls{dsa} termination conditions. Runtime enforcement of these bounds prevents unauthorised processing after approval expiry.

R5: Accounting and Auditability
All security-relevant actions must be recorded in audit logs to support attribution, regulatory audit, and investigation of patient harm or protocol deviations. This requirement is mandated by \gls{hipaa} \S164.312(b) (audit controls for systems processing electronic Protected Health Information) and \gls{gdpr} Art.~30 (records of processing activities), with technical logging and integrity measures justified under Art.~32 (security of processing), and reinforced by ICH GCP E6(R2) requirements for audit trails and sponsor accountability~\cite{ICH_E6R2_Addendum}.

Collectively, R1–R5 define governance obligations that require enforceable system-level controls, motivating the \gls{aaa} architecture described in Section~\ref{sec:system}.

\subsection{Threat Model}
\label{subsec:threat-model}
We consider adversaries who are authenticated participants in a healthcare \gls{fl} federation possessing valid credentials, network access, and the ability to invoke federation APIs, but who may lack valid clinical approval at the time of a specific request. We focus on governance layer threats arising from misalignment between clinical approval and system enforcement.

We identify four primary threat categories that the \gls{aaa} framework must address. These threats represent the core governance violations that can occur when authenticated participants operate outside their authorised boundaries, and do not include threats addressed by complementary mechanisms such as robust aggregation for Byzantine model updates~\cite{blanchard2017machine} or differential privacy for information leakage~\cite{dworkAlgorithmicFoundationsDifferential2014}.

A1: Unauthorised Participation.
Participation without valid institutional approval, after approval expiry, or outside the approved protocol scope. Mitigation: R1, R2, R4.

A2: Privilege Misuse.
An authorised participant attempts actions beyond its approved role (e.g., an evaluator submitting training updates). Mitigation: R3, with R2 constraining actions to approved study context.

A3: Accountability Evasion.
An adversary attempts to influence execution while avoiding traceability. Mitigation: R5 via tamper-evident audit records binding subject identity, action, study identifier, timestamp, and policy version.

A4: Policy Bypass via Incomplete Context.
An adversary attempts to invoke federation actions while omitting required attributes or inducing policy evaluation failures to exploit default permit behaviour. Mitigation: Fail-closed policy evaluation at all enforcement points.\\

Each threat is mitigated through enforcement at defined \gls{fl} lifecycle points: authentication during node activation (R1), fail-closed authorisation at federation approval and per-action execution (R2, R3, R4), and integrity-protected audit events for non-repudiation (R5). Table~\ref{tab:regulatory_mapping} summarises these mappings.

\begin{table}[t]
\centering
\caption{Mapping governance obligations to \gls{fl} enforcement mechanisms}
\label{tab:regulatory_mapping}
\footnotesize
\setlength{\tabcolsep}{3pt}
\renewcommand{\arraystretch}{1.08}
\begin{tabular}{p{1.4cm} p{1.8cm} p{2.0cm} p{2.0cm}}
\toprule
\textbf{Obligation} & \textbf{Regulatory basis} & \textbf{\acrshort{fl} mechanism} & \textbf{FLA\textsuperscript{3} component} \\
\midrule
Institutional identity
& \acrshort{hipaa} \S164.312(a), \S164.308(a)(4); \acrshort{gdpr} Art.~5(2)
& \acrshort{mtls} with institutional credentials
& Flower authentication + \acrshort{pdp} attributes (R1) \\

Institutional approval
& \gls{hra}/REC (UK); EU CTR (NL); ICMR (IN)
& Approval and \acrshort{dsa} status as policy attributes
& \acrshort{pdp} membership validation \\

Purpose limitation
& \acrshort{gdpr} Art.~5(1)(b); \acrshort{dpdpa}
& Study-scoped policies with immutable study IDs
& Study-scoped federation (Section~\ref{sec:multistudy}) + study validation \\

Access control
& \acrshort{hipaa} \S164.312(a) (US); Caldicott Principles (UK)
& Role and action-level enforcement
& \acrshort{pdp} role policies \\

Auditability
& \acrshort{hipaa} \S164.312(b); \acrshort{gdpr} Art.~30
& Integrity protected audit logs
& Cryptographic audit logging with \acrshort{jws} signatures (R5) \\

Temporal validity
& Ethics approval validity periods
& Time-bounded policy conditions with auto-expiry
& \acrshort{pdp} temporal predicates $V_\sigma$, $V_m$ (R4) \\

\bottomrule
\end{tabular}
\end{table}

\section{System Architecture}
\label{sec:system}

\begin{figure}[htbp]
    \centering
    \includegraphics[width=1\linewidth]{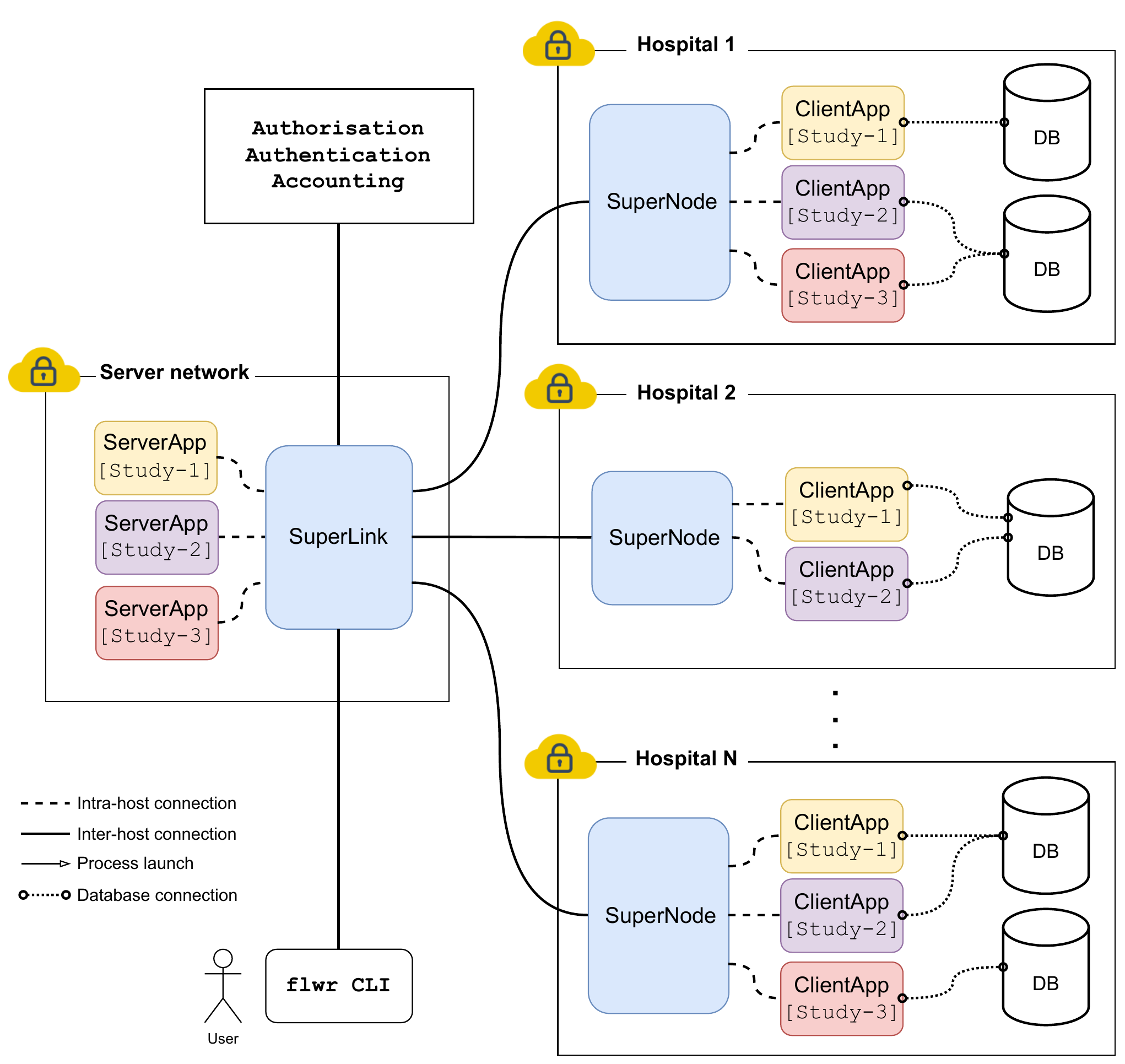}
    \caption{FLA\textsuperscript{3} system architecture. FLA\textsuperscript{3} comprises three layers: central coordination via SuperLink (server), site-local SuperNodes (gateways), and ephemeral ClientApp processes (study-specific execution). The governance layer enforces \gls{aaa} constraints through \gls{xacml}-compliant policy evaluation and cryptographic audit logging.}

    \label{fig:architecture}
\end{figure}

This section presents a governance-aware \gls{fl} architecture designed to satisfy the healthcare governance requirements derived in Section~\ref{sec:governance-threat-model}. The governance mechanisms are framework-agnostic and applicable to any \gls{fl} system with defined orchestration points for policy enforcement. We demonstrate feasibility by extending the Flower runtime~\cite{beutelFlowerFriendlyFederated2022} with study-scoped federation, policy-based access control, and cryptographic accounting for auditable execution.

Fig.~\ref{fig:architecture} illustrates the three-layer architecture: (1) central coordination via SuperLink, (2) site-local SuperNodes, and (3) study-scoped ClientApp processes. This separation enables institutional data sovereignty whilst supporting coordinated multi-study federation under unified governance.

\subsection{FL Runtime}

FLA\textsuperscript{3} adopts Flower's deployment model, which separates coordination, gateway, and execution concerns. The SuperLink acts as the central coordination server, managing study lifecycles, client coordination, and aggregation rounds. Each hospital operates a SuperNode, a site-local gateway that mediates communication between the central server and locally executed application processes.

Model training and evaluation are executed as short-lived ClientApp processes. This separation allows long-running infrastructure components to remain stable and study-specific logic is executed in isolated runs.

Communication between components uses client-initiated \gls{grpc}~\cite{grpc2015} unary calls (single request-response pairs). This design avoids the need for inbound connectivity at hospital sites, which supports the deployment under restrictive institutional network policies (see Section~\ref{sec:deployment}).

\subsection{Multi-Study Federation}
\label{sec:multistudy}

FLA\textsuperscript{3} supports multiple independent studies, each potentially involving a different subset of hospitals. Each study constitutes an independent federation with its own participant set, authorisation policies, and temporal validity constraints. Studies are identified by immutable study identifiers and bounded by explicit validity periods corresponding to ethics approval windows.

This design directly implements study-scoped authorisation (R2). All access control decisions are evaluated within the scope of the requesting study, such that a hospital may act as a participant, observer, or non-member depending on the study context. Temporal constraints (R4) further restrict participation to explicitly authorised study windows.

This study-scoped design reflects real-world research governance, where ethics approvals and data sharing agreements are issued per protocol, and limits the impact of misconfiguration or credential compromise to individual studies.

\subsection{AAA Framework}
\label{sec:aaa}

FLA\textsuperscript{3} implements an explicit \gls{aaa} framework to enforce governance requirements beyond those provided by standard \gls{fl} runtimes.

\subsubsection{Authentication}

We enforce verified institutional participation (R1) through \gls{mtls} with institutional credentials. FLA\textsuperscript{3} builds on Flower's authentication infrastructure, requiring all nodes to present valid certificates issued to recognised healthcare institutions before establishing connections with the federation coordinator.

Authentication binds node identity to institutional credentials but does not grant operational privileges. All subsequent access decisions are evaluated by the authorisation layer.

\subsubsection{Authorisation}

We use a centralised \gls{pdp} based on the \gls{xacml} standard~\cite{OASIS_XACMLv3_2013} to enforce authorisation. \Gls{xacml} provides a declarative language for expressing authorisation policies and a standardised request-response protocol for policy evaluation, enabling separation of policy logic from enforcement mechanisms. The \gls{pdp} is implemented using Luas~\cite{zhangAccessControlImplementation2019}, an \gls{xacml}-compliant policy evaluation engine that we developed previously for resource-constrained environments. Luas operates exclusively on the server side, and Flower server components act as \glspl{pep} that invoke the \gls{pdp} before executing security-sensitive actions.

\paragraph{Request Model}
Let $\mathcal{N}$ denote the set of all registered hospital nodes and $\Sigma$ the set of all approved studies. A request is a tuple $r = (n, a, \sigma, \tau)$ where $n \in \mathcal{N}$ is the requesting node, $a \in \{\texttt{train}, \texttt{evaluate}\}$ is the requested action (model training or evaluation), $\sigma \in \Sigma$ is the study, and $\tau$ is the request timestamp.

\paragraph{Authorisation Predicates}
Study validity enforces temporal constraints (R4):
\begin{equation}
V_\sigma(\sigma, \tau) := \tau < \mathsf{expiry}(\sigma).
\end{equation}

Membership validity enforces study-scoped authorisation (R2, R4):
\begin{equation}
V_m(n, \sigma, \tau) := n \in \mathsf{members}(\sigma) \,\wedge\, \tau < \mathsf{expiry}_m(n, \sigma).
\end{equation}

Role eligibility enforces least-privilege access (R3):
\begin{equation}
V_r(n, a) :=
\begin{cases}
\mathsf{role}(n) = \texttt{p}, & a = \texttt{train}, \\
\mathsf{role}(n) \in \{\texttt{p}, \texttt{o}\}, & a = \texttt{evaluate}.
\end{cases}
\end{equation}

where $\texttt{p} = \texttt{participant}$ and $\texttt{o} = \texttt{observer}$.

\paragraph{Authorisation Decision}
A request $r = (n, a, \sigma, \tau)$ is authorised if:
\begin{equation}
\mathsf{Auth}(r) := V_\sigma(\sigma, \tau) \wedge V_m(n, \sigma, \tau) \wedge V_r(n, a).
\label{eq:auth}
\end{equation}

The \gls{pdp} returns $\texttt{Permit}$ if $\mathsf{Auth}(r)$ evaluates to true. This decision is enforced at each federation lifecycle point: (1) node activation during SuperNode registration, (2) study join request when initiating ClientApp execution, and (3) every communication round before model aggregation. This per-round enforcement ensures temporal validity constraints (e.g., ethics approval expiry) are continuously validated throughout training rather than checked only at provisioning time. Fail-closed semantics apply at all enforcement points: if any predicate evaluates to false or cannot be evaluated (e.g., missing attributes, policy evaluation failure), the request is denied with $\texttt{Deny}$ outcome.

\subsubsection{Accounting and Auditability}

For each authorisation decision, the \gls{pdp} emits a structured audit record encoded in JSON and written to append-only log files. Each record includes the evaluated attributes, the decision outcome, and a canonical JSON payload used for signing. When a signing key is configured, the \gls{pdp} produces a \gls{jws} using the ES256 algorithm~\cite{rfc7515,rfc7518}. Any modification of a signed record invalidates the signature and can be detected using the corresponding public key~\cite{fips186}, providing tamper evidence and non-repudiation to support governance review, compliance auditing, and forensic analysis without reliance on the runtime environment.

\subsection{Federated Aggregation Strategy}
\label{sec:aggregation}

FLA\textsuperscript{3}'s governance mechanisms are agnostic to the choice of federated aggregation algorithm. To demonstrate compatibility with personalised \gls{fl} methods that address statistical heterogeneity in healthcare data, we integrate FedMAP~\cite{zhangFedMAPPersonalisedFederated2025}, a personalised \gls{fl} approach we developed previously.

FedMAP performs local Maximum a Posteriori (MAP) estimation with a learned global prior, enabling sites with heterogeneous data distributions to maintain local accuracy while leveraging shared knowledge from other institutions. At each communication round, participating sites optimise local model parameters under a regulariser that encodes global knowledge, then the server aggregates these local models using posterior-based weighting to update the global prior. Full algorithmic details are provided in our previous work~\cite{zhangFedMAPPersonalisedFederated2025}.

Governance enforcement is integrated directly into the aggregation workflow. At each communication round, the server queries the \gls{pdp} for each registered node $n \in \mathcal{N}$ using request $(n, \texttt{train}, \sigma, \tau)$, where $\sigma$ is the study identifier and $\tau$ is the current timestamp. Only nodes receiving $\texttt{Permit}$ contribute model updates to aggregation.

This integration demonstrates a key architectural principle: governance enforcement operates at the orchestration layer, filtering the participant set before aggregation, rather than within the aggregation algorithm itself. This separation allows the aggregation algorithm to address statistical heterogeneity (in FedMAP's case, through posterior-based weighting) whilst governance policies enforce regulatory compliance. Governance determines who may participate; the aggregation algorithm determines how to combine compliant participants' contributions. 

\subsection{Deployment Considerations}
\label{sec:deployment}

Healthcare deployment environments impose three practical constraints that shape system design. First, clinical network policies typically permit only egress-only connectivity, prohibiting inbound connections; FLA\textsuperscript{3} uses client-initiated \gls{grpc} communication where hospital nodes initiate all connections to the central coordination server, avoiding the need for inbound connectivity or long-lived bidirectional streaming channels. Second, participating institutions exhibit heterogeneous execution environments 
that differ in operating systems, hardware, and approved software stacks, with 
some sites requiring containerised deployment and others preferring native 
installation. This heterogeneity is particularly pronounced in TREs, secure enclaves increasingly 
mandated by UK and EU health data governance frameworks, which impose strict 
controls on approved software, data egress, and network connectivity. 
FLA\textsuperscript{3} accommodates these constraints through pre-built 
container images and native installation procedures, reducing site-specific 
configuration effort. Third, manual code deployment at clinical sites introduces 
operational risk and unintended system interactions; FLA\textsuperscript{3} 
packages study-specific logic into application bundles for runtime deployment, 
removing the requirement for site-side code updates and thereby aligning with 
the software approval processes common in TRE-governed environments.

\section{Experimental Validation}

This section evaluates the proposed FLA\textsuperscript{3} platform from two complementary perspectives. 
Section~\ref{sec:security_validation} validates the correctness and robustness of the access control framework that enforces governance constraints. 
Section~\ref{sec:performance} evaluates the clinical prediction performance of \gls{fl} under realistic multi-centre heterogeneity, comparing individual training, personalised \gls{fl} method, and a centralised reference model.

\subsection{Security Validation}\label{sec:security_validation}

We validated the \gls{xacml}-based authorisation framework implemented within FLA\textsuperscript{3} through systematic testing of all decision paths defined in the XACML \texttt{PolicySet} used by the federation. Test cases were generated using a custom XACML test-case generator developed for this work, which programmatically instantiates both baseline governance requests and adversarial variants designed to exercise policy boundary conditions and potential evasion attempts. The generator and all validation artefacts are released as part of the public code repository accompanying this paper\footnote{\url{https://github.com/bloodcounts/FLAAA/tree/main/pdp/comformance/fl-tests}}.

\subsubsection{Validation Setup}
The validation suite comprises 47 cases organised into two groups.
\emph{Baseline validation} (28 cases) covers the four core policies:
\begin{itemize}
  \item \textbf{P1: StudyApprovalPolicy} — Validates that the requested study identifier corresponds to an approved study with valid temporal bounds (enforces R2, R4).
  \item \textbf{P2: MembershipValidityPolicy} — Verifies institutional membership in the requested study and checks membership has not expired (enforces R1, R2, R4).
  \item \textbf{P3: TrainingAccessPolicy} — Restricts training and aggregation actions to nodes with participant role (enforces R3).
  \item \textbf{P4: EvaluationAccessPolicy} — Permits evaluation actions for both participants and observers (enforces R3).
\end{itemize}
Security and robustness tests (19 cases) target potential bypass vectors, including attribute omission (fail-closed behaviour), malformed datatypes, duplicate attributes, category confusion, input canonicalisation, temporal boundary conditions, and attribute injection.

Each test case specifies the expected authorisation decision (\emph{Permit}, \emph{Deny}, \emph{NotApplicable}, or \emph{Indeterminate}), the policy rule that should match, and the threat category being exercised. \gls{xacml} request documents were evaluated using the Luas \gls{pdp}, and observed decisions were compared against expected outcomes.

\subsubsection{Results}

All 47 validation cases produced the expected authorisation decisions. The policy framework correctly enforced temporal constraints, role separation, and fail-closed behaviour in the presence of missing or malformed attributes. Baseline validation (28 cases) confirmed correct enforcement of study approval (5 cases), membership validity (6 cases), training access restrictions (9 cases), and evaluation permissions (7 cases), along with edge case handling. Security and robustness testing (19 cases) validated protection against attribute omission (6 cases), malformed inputs (3 cases), duplicate attributes (3 cases), category confusion (2 cases), input canonicalisation issues (2 cases), temporal boundary conditions (2 cases), and attribute injection (1 case). Indeterminate decisions were correctly returned for malformed inputs rather than defaulting to permissive outcomes. These results confirm that the governance layer enforces least-privilege access and is robust against common policy evasion strategies, supporting its suitability for deployment in regulated healthcare environments.

\subsection{Clinical Prediction Performance}\label{sec:performance}

\subsubsection{Dataset and Experimental Setup}

Applicability in healthcare was evaluated using data from the INTERVAL randomised controlled trial (ISRCTN24760606)~\cite{diangelantonioEfficiencySafetyVarying2017}, employing routine \gls{fbc} measurements. The prediction task was binary classification of iron deficiency, defined as serum ferritin $<\SI{15}{\micro\gram\per\litre}$, using only routinely available \gls{fbc} parameters \cite{Kreuter_2025, zhangFedMAPPersonalisedFederated2025}.

After standard preprocessing, which excludes unreliable measurements, samples with delays exceeding 36 hours, samples without concurrent FBC and ferritin measurements, and subjects outside eligible height and weight ranges, the dataset comprised \num{54446} samples across 25 donation centres. Centre sizes were highly heterogeneous (coefficient of variation \SI{64.2}{\percent}), reflecting realistic deployment conditions.

We compared three training paradigms:
\begin{enumerate}
  \item Individual training: each centre trained a local model using only its own data.
  \item \gls{fl}: personalised federated training across all centres using FedMAP~\cite{zhangFedMAPPersonalisedFederated2025}.
  \item Centralised training: pooled training across all centres, used as a reference representing an upper bound that is infeasible in practice due to governance constraints.
\end{enumerate}

All models shared the same architecture (two-layer multilayer perceptron with 64 and 32 hidden units).

\subsubsection{Results}
\begin{figure}[t]
\centering
\includegraphics[width=\linewidth]{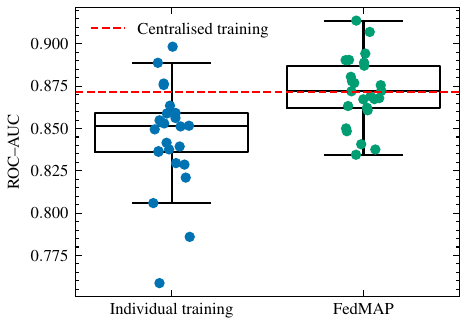}
\caption{Per-centre \gls{rocauc} distributions for individual training and FedMAP. 
Each point corresponds to a donation centre. 
Boxplots summarise the distribution across centres with whiskers extending to 1.5 interquartile ranges. 
The red dashed line indicates centralised reference model performance. 
\glsexplain{rocauc}; higher values indicate better predictive performance.}
\label{fig:boxplot}
\end{figure}

\begin{figure}[t]
\centering
\includegraphics[width=\linewidth]{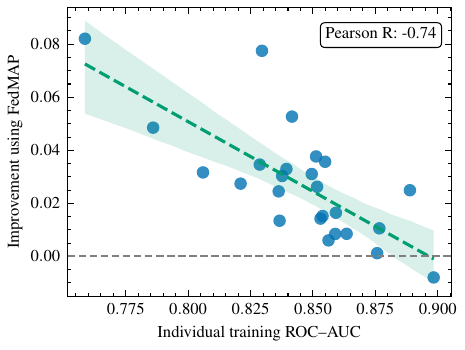}
\caption{Per-centre performance gain from federation (FedMAP minus individual \gls{rocauc}) plotted against baseline individual performance. Centres with lower baseline performance benefit most from federation, indicating mitigation of inter-centre heterogeneity.
\glsexplain{rocauc}.}
\label{fig:gain}
\end{figure}

\begin{figure}[t]
\centering
\includegraphics[width=\linewidth]{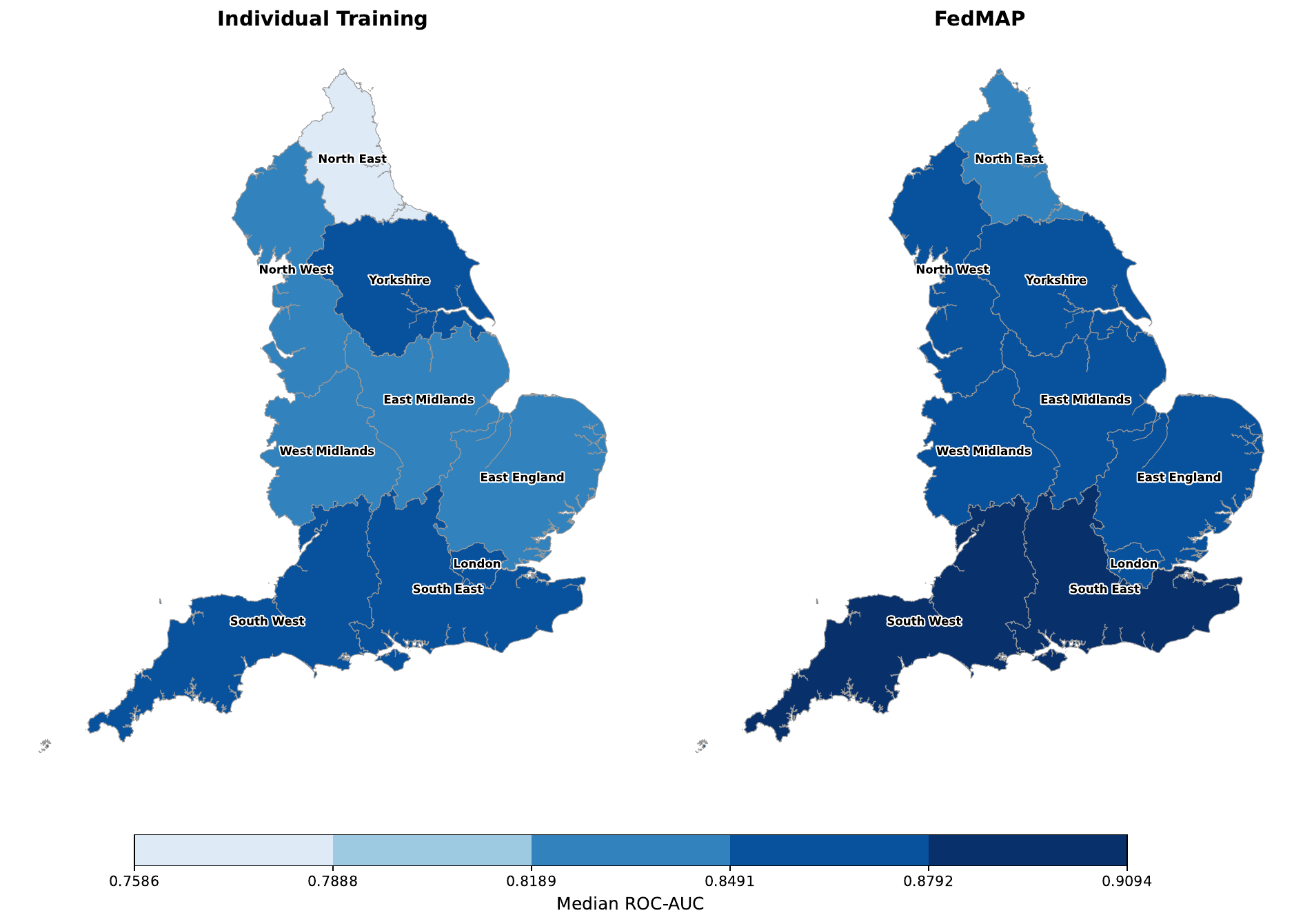}
\caption{Regional median \gls{rocauc} for individual training (left) and FedMAP (right). Centres were aggregated by region using the median \gls{rocauc} to illustrate geographic consistency of performance. Identical colour scales are used across both maps.}
\label{fig:map}
\end{figure}

Across centres, FedMAP achieved a mean \gls{rocauc} of 0.872, compared with 0.845 for individual training, corresponding to an average improvement of 0.027 \gls{rocauc} (95\% CI: [0.019, 0.036]) (Fig.~\ref{fig:boxplot}). This improvement was statistically significant (paired Wilcoxon signed-rank test: statistic $=3.00$, $p<0.001$; Cohen’s $d=1.28$).
The centralised reference model achieved a mean \gls{rocauc} of 0.872, indicating that federated training achieved performance comparable to centralised training while respecting governance constraints.

Of the 25 centres, 24 (96\%) improved under federated training (Fig.~\ref{fig:gain}). The single centre exhibiting a marginal decrease had the highest individual baseline performance (\gls{rocauc} $=0.898$).
There was a strong negative correlation ($r=-0.74$) between baseline individual performance and federated gain, indicating that centres with poorer baseline performance benefited most from collaboration. Federated training also reduced inter-centre variability, with the standard deviation of \gls{rocauc} decreasing from 0.029 to 0.020.

To evaluate whether governance enforced federation improves geographic equity in model performance, a regulatory expectation for multi-centre healthcare AI deployments, we analysed regional variability across NHS England regions. Using individual training per centre, regional median \glspl{rocauc} ranged from 0.759 to 0.876 with a range of 0.117. Federated training reduced this range to 0.844 to 0.909 with a range of 0.065, demonstrating improved geographic consistency. 
Fig.~\ref{fig:map} shows median regional \gls{rocauc} across NHS England regions, illustrating the reduced variability when using \gls{fl}.

\section{Discussion}
\label{sec:discussion}

This work addresses a persistent gap between \gls{fl} research and deployable healthcare systems: the absence of runtime-enforceable governance mechanisms within \gls{fl} orchestration. While prior studies have established the technical feasibility of \gls{fl} for healthcare applications, systematic reviews consistently find that the overwhelming majority implicitly assume trusted participants and idealised operating conditions~\cite{teoFederatedMachineLearning2024,zhangRecentMethodologicalAdvances2024}. We demonstrate that cross-jurisdictional healthcare regulations can be systematically analysed to derive enforceable governance requirements, and that these requirements can be integrated directly into FLA\textsuperscript{3} through policy-driven access control and cryptographic accountability.

The proposed \gls{aaa} framework provides several technical contributions that distinguish it from existing approaches. First, the \gls{xacml}-based authorisation layer operates with fail-closed semantics: missing attributes, policy evaluation failures, or ambiguous decisions result in denial rather than permissive defaults. This design choice directly addresses the governance threat model (Section~\ref{sec:governance-threat-model}), wherein authenticated participants may lack valid clinical approval at the time of a specific request. Second, cryptographically signed audit records provide tamper-evident accountability that supports regulatory audit requirements independently of the runtime environment. Third, the study-scoped federation model aligns system behaviour with clinical research practice, where ethics approvals and \glspl{dsa} are issued per protocol and do not extend across studies \cite{ICH_E6R3_2025, WHO_Ethics_Review_2011}.

A central finding of this work is that governance enforcement does not degrade predictive performance. Crucially, the \gls{aaa} framework provides the legal and operational governance mechanisms needed to support such federation across heterogeneous sites, which would otherwise have been prohibited by regulatory barriers. Across centres, federated training achieved performance comparable to centralised reference models while strictly enforcing governance requirements, and the majority of institutions benefited relative to individual training. Improvements were most pronounced for centres with lower baseline performance, and federation reduced inter-centre and inter-regional variability. These results suggest that \gls{fl} can mitigate institutional and geographic disparities in model performance while enforcing governance compliance, a consideration of direct relevance to health equity objectives \cite{hasanzadehBiasRecognitionMitigation2025}. Similarly, the reduction in regional variability, from a range of 0.117 to 0.065 in median \gls{rocauc}, suggests that governance-compliant \gls{fl} may help address geographic disparities in predictive model performance.

The operational deployment across five international healthcare institutions provides evidence of practical feasibility under realistic constraints. Participating institutions operate under distinct regulatory frameworks (\gls{gdpr}, \gls{dpdpa}, \gls{ecowas}) and exhibit heterogeneous network policies. The architecture accommodates egress-only network configurations common in hospital environments \cite{NHS_HSCN_Security_2023,mullerNavigatingRealWorld2025} and distributes governance policies alongside study logic through Flower's \gls{fab} mechanism, thereby reducing site-specific configuration requirements.

It is important to acknowledge several limitations. First, the clinical evaluation employed simulated federation of INTERVAL data to enable controlled comparison; fully live end-to-end training across deployed institutions remains subject to ongoing governance approvals. Second, the governance framework prevents unauthorised participation and enforces approved roles through authentication and authorisation controls. Byzantine adversaries holding valid credentials who submit malicious model updates require complementary algorithmic defenses (e.g., robust aggregation~\cite{blanchard2017machine}), as \gls{aaa} controls which entity participates but does not validate the integrity of model updates during training. Third, \gls{xacml} policies are manually authored from regulatory analysis, introducing potential for specification errors; formal verification of policy correctness and tools for policy conflict detection represent areas for future investigation. Fourth, although the framework is designed to integrate with algorithmic privacy mechanisms such as differential privacy and secure aggregation, this integration has not been empirically validated in the current deployment.

The governance layer is architecturally separable from the learning algorithm. Policy enforcement operates at the orchestration level, filtering the participant set prior to aggregation, whilst FedMAP addresses statistical heterogeneity through its learned prior mechanism. This separation enables healthcare institutions to adopt alternative aggregation strategies appropriate to their data characteristics and risk tolerance whilst retaining the governance controls proposed in this work. The \gls{aaa} framework can be integrated with any \gls{fl} approach requiring institutional accountability and protocol compliance in deployed settings.

More broadly, this work establishes that compliance and utility need not be in tension: governance constraints can be satisfied without sacrificing the collaborative benefits that motivate \gls{fl} adoption.

\section{Conclusion and Future Work}
\label{sec:conclusion-future}

This paper presented FLA\textsuperscript{3}, a governance-aware \gls{fl} platform that integrates enforceable access control, temporal validity, and cryptographic accountability directly into \gls{fl} orchestration as runtime system properties.

Security validation confirms that the \gls{xacml}-based policy framework correctly enforces all specified governance constraints under both standard and adversarial attribute conditions, with fail-closed behaviour preventing policy bypass through attribute omission or malformation. Clinical evaluation using simulated federation of data from 25 blood donor centres demonstrates that governance-compliant federated training can achieve predictive performance comparable to a centralised reference model while reducing inter-centre and inter-regional variability. Separately, operational deployment of the platform infrastructure across five international healthcare institutions demonstrates practical feasibility under realistic network and governance constraints.

These results establish that governance enforcement is achievable within \gls{fl} systems without compromising learning utility, addressing a practical concern that has limited healthcare adoption of collaborative learning approaches \cite{teoFederatedMachineLearning2024}. The modular architecture enables composition with complementary privacy-enhancing techniques and alternative aggregation strategies, providing flexibility for institutions with varying risk profiles and data characteristics.

Future work will focus on three directions. First, completing fully live end-to-end federated training workflows across deployed institutions as governance approvals permit, thereby moving beyond simulated federation to validated multi-institutional learning. Second, empirically evaluating composition with differential privacy and secure aggregation mechanisms to provide layered privacy protection alongside governance enforcement. Third, developing policy authoring tools and formal verification methods to reduce the risk of specification errors and support institutions in translating regulatory requirements into correct \gls{xacml} policies. These extensions aim to reduce adoption barriers for large-scale, multi-jurisdictional healthcare collaborations where both regulatory compliance and predictive performance constitute essential requirements.

\appendices

\section{BloodCounts! Consortium Members}

\noindent
Martijn Schut$^{1}$, Folkert Asselbergs$^{1}$, Sujoy Kar$^{2}$, Suthesh Sivapalaratnam$^{3}$, Sophie Williams$^{3}$, Mickey Koh$^{4}$, Yvonne Henskens$^{5}$, Norbert C.J. de Wit$^{5}$, Umberto D'Alessandro$^{6}$, Bubacarr Bah$^{6}$, Ousman Secka$^{6}$, Parashkev Nachev$^{7}$, Rajeev Gupta$^{7}$, Sara Trompeter$^{7}$, Nancy Boeckx$^{8}$, Christine van Laer$^{8}$, Gordon A. Awandare$^{9}$, Kwabena Sarpong$^{9}$, Lucas Amenga-Etego$^{9}$, Mathie Leers$^{10}$, Mirelle Huijskens$^{10}$, Samuel McDermott$^{11}$, Willem H. Ouwehand$^{12}$, James Rudd$^{13}$, Carola-Bibiane Sch{\"o}nlieb$^{11}$, Nicholas Gleadall$^{12,14,15}$, and Michael Roberts$^{11,13}$.

\vspace{1em}

\noindent
$^{1}$ Amsterdam University Medical Centre, Amsterdam, Netherlands.\\
$^{2}$ Apollo Hospitals, Chennai, India.\\
$^{3}$ Barts Health NHS Trust, London, United Kingdom.\\
$^{4}$ Health Services Authority, Singapore.\\
$^{5}$ Maastricht University Medical Centre, Maastricht, Netherlands.\\
$^{6}$ MRC The Gambia Unit, Banjul, The Gambia.\\
$^{7}$ University College London Hospitals, London, United Kingdom.\\
$^{8}$ University Hospitals Leuven, Leuven, Belgium.\\
$^{9}$ West African Centre for Cell Biology of Infectious Pathogens, Accra, Ghana.\\
$^{10}$ Zuyderland Medical Center, Zuyderland, Netherlands.\\
$^{11}$ Department of Applied Mathematics and Theoretical Physics, University of Cambridge, UK.\\
$^{12}$ NHS Blood and Transplant, Cambridge, UK.\\
$^{13}$ Department of Medicine, University of Cambridge, UK.\\
$^{14}$ Victor Phillip Dahdaleh Heart and Lung Research Institute, University of Cambridge, UK.\\
$^{15}$ Department of Haematology, University of Cambridge, UK.

\section*{Acknowledgment}
F.~Zhang, D.~Kreuter, S.~Sivapalaratnam, J.~Taylor, C.-B.~Schönlieb, N.~S.~Gleadall, and M.~Roberts have received support from the Trinity Challenge grant awarded to establish the BloodCounts! consortium, along with NIHR UCLH Biomedical Research Centre, the NIHR Cambridge Biomedical Research Centre, National Health Service Blood and Transplant (NHSBT) and the Medical Research Council. 
D.~Kreuter and J.~Taylor receive support from MRC GAP Fund (UKRI/814). B.~Butler acknowledges support of Taighde Éireann – Research Ireland under Grant number 13/RC/2077\_P2 CONNECT.
N.~S.~Gleadall has been supported by NHSBT grants 1701-GEN; 20-01-GEN; G120400. 
C.-B.~Schönlieb acknowledges support from the EPSRC programme grant in `The Mathematics of Deep Learning' (EP/L015684), Cantab Capital Institute for the Mathematics of Information, the Philip Leverhulme Prize, the Royal Society Wolfson Fellowship, the EPSRC grants EP/S026045/1, EP/T003553/1, EP/N014588/1, and EP/T017961/1, the Wellcome Innovator Award RG98755 and the Alan Turing Institute. 
M.~Roberts is additionally supported by the British Heart Foundation (TA/F/20/210001). 
(The views expressed are those of the authors and not necessarily those of the NIHR or the Department of Health and Social Care).

Participants in the INTERVAL randomised controlled trial were recruited with the active collaboration of NHS Blood and Transplant England (\url{https://www.nhsbt.nhs.uk}), which has supported field work and other elements of the trial. DNA extraction and genotyping were co-funded by the National Institute for Health and Care Research (NIHR), the NIHR BioResource (\url{http://bioresource.nihr.ac.uk}) and the NIHR Cambridge Biomedical Research Centre (BRC-1215-20014)\footnote{The views expressed are those of the authors and not necessarily those of the NIHR or the Department of Health and Social Care.}.

\medskip

The academic coordinating centre for INTERVAL was supported by core funding from the: NIHR Blood and Transplant Research Unit (BTRU) in Donor Health and Genomics (NIHR BTRU-2014-10024), NIHR BTRU in Donor Health and Behaviour (NIHR203337), UK Medical Research Council (MR/L003120/1), British Heart Foundation (SP/09/002; RG/13/13/30194; RG/18/13/33946), NIHR Cambridge BRC (BRC-1215-20014; NIHR203312), and by Health Data Research UK, which is funded by the UK Medical Research Council, Engineering and Physical Sciences Research Council, Economic and Social Research Council, Department of Health and Social Care (England), Chief Scientist Office of the Scottish Government Health and Social Care Directorates, Health and Social Care Research and Development Division (Welsh Government), Public Health Agency (Northern Ireland), British Heart Foundation and Wellcome.

\medskip

A complete list of the investigators and contributors to the INTERVAL trial is provided in reference~\cite{diangelantonioEfficiencySafetyVarying2017}. The academic coordinating centre for INTERVAL would like to thank blood donor centre staff and blood donors for participating in the INTERVAL trial.

\section*{Ethics Approval and Consent to Participate}
The INTERVAL trial is compiled into the Blood Donors Studies BioResource (BDSB) with Research Ethics Committee (REC) reference 20/EE/0115. All participants provided informed consent for their data to be used in research studies.

\section*{Data Availability}
Access to BDSB data can be applied for from the Blood Donors Studies Data Access Committee. The FLA\textsuperscript{3} platform implementation, including XACML policy templates and security validation test suites, is publicly available at \url{https://github.com/bloodcounts/FLAAA}.

\section*{Conflict of Interest}
J. Fernandez-Marques and N. Lane are employed by Flower Labs, 
the organisation that develops the Flower federated learning 
framework upon which the FLA\textsuperscript{3} platform is built. G. Verghese is employed by PharosAI.
The remaining authors declare no conflicts of interest relevant 
to this work.

\section*{References}

\bibliographystyle{IEEEtran}

\bibliography{references}

\end{document}

%% file: acronyms.tex
\newacronym{fl}{FL}{federated learning}
\newacronym{aaa}{AAA}{authentication, authorisation, and accounting}
\newacronym{ai}{AI}{artificial intelligence}
\newacronym{cli}{CLI}{command-line interface}

\newacronym{hipaa}{HIPAA}{Health Insurance Portability and Accountability Act}
\newacronym{gdpr}{GDPR}{General Data Protection Regulation}
\newacronym{dpdpa}{DPDPA}{Digital Personal Data Protection Act}
\newacronym{ecowas}{ECOWAS}{Economic Community of West African States}
\newacronym{xacml}{XACML}{eXtensible Access Control Markup Language}

\newacronym{pdp}{PDP}{policy decision point}
\newacronym{pep}{PEP}{policy enforcement point}
\newacronym{rbac}{RBAC}{role-based access control}
\newacronym{abac}{ABAC}{attribute-based access control}

\newacronym{mtls}{mTLS}{mutual transport layer security}
\newacronym{tls}{TLS}{transport layer security}
\newacronym{jws}{JWS}{JSON web signature}
\newacronym{grpc}{gRPC}{gRPC remote procedure call}
\newacronym{fab}{FAB}{Flower App Bundle}
\newacronym{tee}{TEE}{trusted execution environment}

\newacronym{map}{MAP}{maximum a posteriori}
\newacronym{icnn}{ICNN}{input convex neural network}
\newacronym{noniid}{non-IID}{non-independent and identically distributed}
\newacronym{rocauc}{ROC--AUC}{receiver operating characteristic area under the curve}

\newacronym{fbc}{FBC}{full blood count}
\newacronym{dsa}{DSA}{data sharing agreement}

\newacronym{nhs}{NHS}{National Health Service}
\newacronym{nihr}{NIHR}{National Institute for Health Research}
\newacronym{nhsbt}{NHSBT}{NHS Blood and Transplant}
\newacronym{mrc}{MRC}{Medical Research Council}
\newacronym{epsrc}{EPSRC}{Engineering and Physical Sciences Research Council}
\newacronym{ccmo}{CCMO}{Central Committee on Research Involving Human Subjects}
\newacronym{hra}{HRA}{UK Health Research Authority}